\documentclass{article}

\usepackage{arxiv}

\usepackage[utf8]{inputenc} 
\usepackage[T1]{fontenc}    
\usepackage{hyperref}       
\usepackage{url}            
\usepackage{booktabs}       
\usepackage{amsfonts}       
\usepackage{nicefrac}       
\usepackage{microtype}      
\usepackage{lipsum}		
\usepackage{graphicx}
\usepackage{natbib}
\usepackage{doi}
\usepackage{algorithm}

\usepackage{multirow}
\usepackage{array,booktabs,xcolor}
\usepackage{amsmath}
\usepackage{caption}
\usepackage{url}
\usepackage{dsfont}

\UseRawInputEncoding 

\title{An automatic procedure to determine groups of  nonparametric regression curves}

\author{ \href{https://orcid.org/0000-0002-3224-8858}{\includegraphics[scale=0.06]{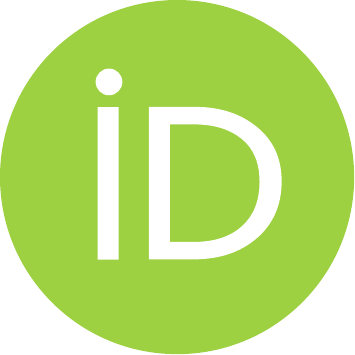}\hspace{1mm}Nora M. Villanueva} \\
	Department of Statistics and Operational Research\\
	SiDOR Group and CINBIO\\
	University of Vigo, Vigo, Spain \\
	\texttt{nmvillanueva@uvigo.es} \\
	\And
	\href{https://orcid.org/0000-0003-4284-6509}{\includegraphics[scale=0.06]{orcid.pdf}\hspace{1mm}Marta Sestelo} \\
	Department of Statistics and Operational Research\\
	SiDOR Group and CINBIO\\
	University of Vigo, Vigo, Spain \\
	\texttt{sestelo@uvigo.es} \\
	\And
	Celestino Ord\'o\~nez \\
	Department of Mining Exploitation and Prospecting\\
	University of Oviedo, Oviedo,  Spain\\
	\texttt{rordonezcelestino@uniovi.es} \\
	\And
	Javier Roca-Pardi\~nas \\
	Department of Statistics and Operational Research\\
	SiDOR Group and CINBIO\\
	University of Vigo, Vigo, Spain \\
	\texttt{roca@uvigo.es} }

\renewcommand{\shorttitle}{An automatic procedure to determine groups of  nonparametric regression curves}

\hypersetup{
pdftitle={An automatic procedure to determine groups of  nonparametric regression curves},
pdfsubject={stat.ME},
pdfauthor={Nora M. Villanueva et al.},
pdfkeywords={First keyword, Second keyword, More},
}

\begin{document}
\maketitle

\begin{abstract}
In many situations it could be interesting to ascertain whether nonparametric regression curves can be grouped, especially when confronted with a considerable number of curves.  The proposed testing procedure allows to determine groups with an automatic selection of their number.  A simulation study is presented in order to investigate the finite sample properties of the proposed methods when compared to existing alternative procedures.   Finally,  the applicability of the procedure to study the geometry of a tunnel by analysing a set of cross-sections is demonstrated. The results obtained show the existence of some heterogeneity in the tunnel geometry.
\end{abstract}


\section{Introduction}
\label{noramv:intro} 
One of the main goals of statistical modelling is to understand the dependence of a response variable, $Y$, with respect to another explanatory variable, $X$. This type of dependence can be studied through nonparametric regression models, where the relationship between $Y$ and $X$ is modelled without specifying in advance the function that links them. Within this framework, the study of the regression curves can be useful in the comparison of two or more groups, which is an important problem associated with statistical inference.  In particular, the topic of hypothesis testing the equality of mean functions has been widely investigated in the literature, see, for instance, the review that  \cite{crujeiras2013} offers about this topic. Relevant papers on this topic are \cite{hall1990,king91,Delgado,kulasekera95,bowman95,dette2001,Pardo,Srihera:2010:NCR:1837522.1837642}, among others.  Furthermore, in order to compare the values of a response variable across several groups in the presence of a covariate effect, nonparametric analysis of covariance  or  factor-by-curve interaction test can be used.  \cite{bowman95} generalized the one-way analysis of variance test  to the nonparametric regression setting, and  \cite{dette2001} proposed to use Young and Bowman's test also in the situation of a heteroscedastic error. In addition,  \cite{park2008} developed a  SiZer tool based on an analysis of variance  type test statistic  that is capable of  comparing  multiple curves based on the residuals. The evolution of this procedure is based on the comparison using  the original regression curves \citep{park2014}.  More recently,  the possibility of comparing  curves as well as their derivatives has  been proposed by \cite{sesteloJRSSC} using factor-by-curve interactions in a nonparametric framework.    

When the null hypothesis of equality of curves is rejected, leading to the clear conclusion that at least one  curve is different,  it can be interesting to ascertain whether groups of curves can exist or, by contrast, that all these curves are different from each other. In this setting, one na\"ive approach would be to perform pairwise comparisons. Following the ideas of  \cite{Rosenblatt1975,cao,hardle}, an alternative test to check the null hypothesis of equality of curves obtained from a pairwise comparison of the estimators of the regression functions was proposed by \cite{dette2001}. A similar statistic was considered by \cite{king91}. However, such approaches lead to difficult interpretation of results because as  the number of curves increases so does the number of comparisons. For example, considering 50 curves, the number of  all pairwise comparisons that need to be conducted is 1225. One could make it but without the possibility of determining groups with similar regression curves.

With this focus but in time-to-event framework, \cite{villanueva2019} have described a new procedure to determine groups in multiple survival curves. Some approaches have been developed also in  functional data context \citep{abraham03,garciaescudero05,tarpey07} and in longitudinal one  \citep{vogt2016,vogt2020}. However, to the best of our knowledge the problem of determining  groups of regression functions in a standard nonparametric framework (cross-sectional data) has not been considered explicitly in the literature.

Based on the above, and taking into account situations with a considerable  number of curves,  we introduce an approach  that allows us to  group multiple regression curves. Briefly, our procedure is described as follows.  Firstly, the $J$ regression curves are estimated  by  kernel smoothers. Secondly, given a number of  $K$ groups,  the optimal    possible assignment  of  $J$ curves into $K$ groups is  chosen  by means of  a heuristic algorithm. Finally,  the optimal number of groups  is determined using  an automatic bootstrap-based testing procedure.  

The proposed methodology is used to  study the geometry of a tunnel by analysing a set of cross-sections along it. These sections were obtained by adjusting a surface to a point cloud measured with a terrestrial laser scanning. The existence of different groups of cross-sections indicates that the tunnel has a heterogeneous geometry. This heterogeneity could be due to bad construction or to deformations in the tunnel.

The remainder of this paper is organized as follows. In Section \ref{noramv:estimation}, the notation and the methodological background are explained.  In Section \ref{noramv:simulationest}, the performance of  our and existing procedures is shown through  simulation studies. The results of the analysis of a  real dataset are provided in Section \ref{noramv:app}. Finally,  the main conclusions of this work are exposed in Section \ref{noramv:conclusion}.

\section{Methodology}
\label{noramv:methodology}

\subsection{Notation and technical details}
\label{noramv:estimation}

 Let $(X_j,Y_j)$ be $J$ independent random vectors, and assume that they satisfy the following nonparametric regression models,  for $j=1,\ldots,J$,
\begin{equation} 
Y_j=m_j(X_j) + \varepsilon_j
\label{noramv:eq1}  
\end{equation}
where $m_j$ is a nonparametric smooth function and $\varepsilon_j$ is the regression error, which is assumed independent of the covariate $X_j$. Note that, by construction,  $E(\varepsilon_j) = 0$ and $Var(\varepsilon_j)=\sigma_j^2$, which means that $m_j(X_j)$ represents the conditional mean, while $\sigma_j^2$ equals the residuals variance, i.e.,  $\sigma_j^2 = Var(Y_j  - m_j(X_j))$. Throughout this paper, we will not require any  assumptions on the error distributions.

Several approaches can be used to estimated the regression models in (\ref {noramv:eq1}), such as, methods based on regression splines \citep{deboor}, Bayesian approaches \citep{Lang04} or local  polynomial kernel smoothers \citep{NoRefWorks:7,GVK19282144X}.  In this work, local  linear kernel smoothers, as implemented in the npregfast R package \citep{npregfast,sestelojss}, are used. Explicitly, given $J$  independent random samples, say 

\[ \left\{{ S_1=\left\{ { \left( {X_{i1}, Y_{i1}} \right)}\right\}_{i=1}^{n_1},\ldots, S_J=\left\{ { \left( {X_{iJ}, Y_{iJ}} \right)}\right\}_{i=1}^{n_J} }\right\} \]
where the random variables $(X_{1j}, Y_{1j}),\ldots,(X_{n_jj}, Y_{n_jj})$ are i.i.d.  for each $j=1,\ldots,J$ and with a total sample size $n=\sum_{j=1}^J n_j$, the local linear kernel estimator 
 
\begin{equation}
\hat m_j(x)=\Psi(x,S_j,h_j) 
\label{jrp:eq2}  
\end{equation}
at a location $x$ is given by $\hat m_j(x)= \hat {\alpha} _{0j} (x)$, where $\hat{\alpha}_{0j}(x)$ is the first element of the vector $ (\hat \alpha_{0j}(x),\hat \alpha_{1j}(x))$ which is the minimiser  of 

\begin{equation}
\sum_{i = 1}^{n_j} \left \{ {Y_{ij}-\alpha_{0j} (x)-\alpha_{1j} (x) (X_{ij}-x)} \right\}^2\cdot \kappa\left( {{\frac{{X_{ij} - 
x}}{h_j}}} \right),
\label{h0}
\end{equation}
\noindent
where $\kappa$ denotes a kernel function (normally, a symmetric density), and $h_j>0$ is the smoothing parameter or bandwidth. Taking into account that nonparametric estimates depend heavily on the bandwidth $h_j$, various  methods for an optimal selection have been suggested, such as  Generalised Cross-Validation \citep{golub} or plug-in methods, see e.g., \cite{1995}.  See \cite{NoRefWorks:7} for a good overview of this topic. However, optimal bandwidth selection is still a challenging problem. As a practical solution, and based on the simulation results, the bandwidths $h_j$ can be selected automatically  by minimising the  following cross-validation criterion
\begin{equation}
CV(h_j) = {\sum\nolimits_{i = 1}^{n_j} {\left( { {Y}_{ij} - \hat m_j^{-i} (X_{ij}) } \right)}} ^2,
\label{cv}
\end{equation}
where $\hat m_j^{-i} (X_{ij})$ indicates  the fit at $X_{ij}$ leaving out the $i$-th data point based on the smoothing parameter $h_j$.

Additionally, it is well-known that bootstrap resampling techniques are time-consuming processes because the estimation of the model is carried out many times. Moreover, the use of cross-validation for selecting the bandwidth implies a high computational cost because it is necessary to repeat the estimation procedure several times to select the optimal estimate. Accordingly, to apply some computational acceleration technique is fundamental to ensure that the problem can be addressed adequately in practical situations. Thus, in this paper we use binning techniques to speed up the process. A detailed explanation of this technique can be found in  \cite{NoRefWorks:12}.


\subsection{The algorithm for determining groups}
\label{noramv:clustering}

As pointed out in the Introduction, several nonparametric methods have been proposed in the literature in order to test the equality of regression curves, i.e., to test the null hypothesis $H_0: m_1= \ldots = m_J$. If the test is statistical significant and this hypothesis is thus rejected, then determining groups of regression curves becomes of interest, that is, assessing if  the levels  $\{1,\ldots,J\}$ can be grouped in  $K $ groups $G_1,\ldots,G_K$ with $K<J$ such that for each $k=1,\ldots,K$,  $m_i=m_j$ for all  $i,j \in G_k$.  Note that $(G_1, \ldots, G_K)$ must be a partition of $\{1, \ldots, J\}$, and therefore must satisfy  
$G_1 \cup \ldots \cup G_K= \{1,\ldots,J\}$ and  $C_i \cap G_j= \emptyset$, for all $i \neq j \in \{1,\ldots,K\}$.

Let $(X_{ij}, Y_{ij})$, $i = 1, \ldots, n_j$, be an  i.i.d. sample from the distribution of $(X_{j}, Y_{j})$, for each $j= 1, \ldots, J$, and with the total sample size $n=\sum_{j=1}^J n_j$, we propose a procedure to test,  for a given a number $K$,  the null hypothesis $H_0(K)$ that at least one partition exists $(G_1,\ldots,G_K)$ so that all the conditions  above are satisfied. The alternative hypothesis $H_1(K)$ is that for any $(G_1,\ldots,G_K)$, exists at least a group $G_k$  in which  $m_i\neq m_j $ for some $i,j \in G_k$. 

The testing procedure is based on the $J$-dimensional process
\[ 
{\bf{\widehat{V}}}(z) = ( \widehat V_1(z), \widehat V_2(z), \ldots, \widehat V_J(z) )^t,
\]
where, for $j = 1, \ldots, J$,
\[ 
\widehat V_j(z) = \sum_{k = 1}^{K}[\widehat m_j(z) - \widehat{f}_k(z)] \hspace{0.1cm} I_{\{j \in G_k\}}
\]
and $\widehat f_k$  is the pooled local linear kernel estimate based on the combined $G_k$-partition sample, i.e., 
\[ 
\widehat{f}_k(z) = \Psi(z,S_k,h_k) 
\]
where
\[
S_k = \sum_{j \in G_k} S_j.
\]

The following test statistics were considered in order to test $H_0(K)$: a Cram\'er-von Mises type test statistic 
\[
D_{CM} = \min \limits_{G_1,\ldots,G_K} \quad \sum \limits_{j = 1}^{J}   \int_{R_X} {\widehat{V}^2_j(z)} dz 
\]
and a modification of it based on the $L_1$ norm proposed in the Kolmogorov-Smirnov test statistic
\[
D_{KS} = \min \limits_{G_1,\ldots,G_K} \quad \sum \limits_{j = 1}^{J}   \int_{R_X} {|\widehat{V}_j(z)|} dz 
\]


In order to solve the minimisation problem in each test statistic, all the different assignments of the $J$ curves into $K$ groups have to be evaluated. Because of the large number of calculations, this method is feasible only for small numbers of $J$ and $K$. Obviously, when confronted with  a large number of  $J$ curves, the procedure requires an excessively high computational cost. To be more specific, we deal in our  study  of tunnel cross-sections with $J = 66$ curves and $K = 5$ groups and, taking into account  the formula of \cite{Jain88},  the total number of different assignments is 1e44. This combinatorial explosion implies that the problem becomes intractable. Therefore, heuristic algorithms would be really appropriate to tackle much bigger problems.  One the one hand, in the case of $D_{CM}$ test statistic which is defined in terms of the $L_2$-distance, we propose the use of the $K$-means  \citep{MacQueen67}. On the other hand, for the $D_{KS}$ test statistic defined in terms of the $L_1$-norm, the $K$-medians \citep{MacQueen67,books/wi/KaufmanR90} would be more suitable. In both cases, the carried out procedure is equivalent: the  regression functions $m_j$ ($j= 1, \ldots, J$) have to be estimated in a common grid of size $Q$ leading to a matrix of ($J$ x $Q$) dimension, where each row corresponds with the estimates of the $j$ curve in the $Q$ positions of the grid. Then, this matrix will be the input of both heuristic methods, $K$-means and $K$-medians, and from these the ``best'' partition ($G_1,\ldots,G_K$)  is obtained.

Finally, the decision rule based on $D$ consists of rejecting the null hypothesis if $D$ is larger than $(1-\alpha)$-percentile obtained under  the null hypothesis.  To approximate the distributions of the test statistics, resampling methods such as the bootstrap introduced by \cite{NoRefWorks:8} (see also \cite{NoRefWorks:9,NoRefWorks:18,NoRefWorks:11}) can be applied instead. Here we use the wild bootstrap \citep{citeulike:785121,liu:1988,citeulike:991599} because this method is valid both for homoscedastic and for heteroscedastic models where the variance of the error is a function of the covariate. 

The testing procedure used  requires the following steps:
          
\noindent  \textbf{Step 1.} Using the original sample, for $j = 1, \ldots, J$ and $i = 1, \ldots, n_j$, estimate in a common grid the regression functions $m_j$, using each sample separately. Then, applying the proposed algorithms, obtain the ``best'' partition  ($G_1,\ldots,G_K$) and with it, obtain the estimated curves $\widehat f_k$  using a pooled local linear kernel estimator based on the combined partition samples.

\noindent \textbf{Step 2.}  Obtain the $D$ value as explained before, and the null errors under the $H_0(k)$ as
\[
\widehat \varepsilon_{ij}=\sum _{k=1}^K  \left( {Y_{ij}- \widehat{f}_k(X_{ij})}\right) I_{\{j \in G_k\}}. 
\]

\noindent \textbf{Step 3.} Draw bootstrap samples as follows, for $b = 1, \ldots ,B$, and for each $j \in G_k$, \\ draw 
$\left\{ { \left( {X_{i1}, Y_{i1}^{\ast b}} \right)}\right\}_{i=1}^{n_1},\ldots,\left\{ { \left( {X_{iJ}, Y_{iJ}^{\ast b}} \right)}\right\}_{i=1}^{n_J}$
where  
\[
Y_{ij}^{\ast b}=  \sum _{k=1}^K  \hat f_k(X_{ij}) I_{\{j \in G_k\}}
    +\varepsilon_{ij}^{\ast b}
\]
and 
\[\varepsilon_{ij}^{\ast b}= \widehat \varepsilon_{ij} W_i\]
being $W_1, \ldots, W_n$  an i.i.d. random variables with mass $(5+\sqrt{5})/{10}$ and $(5-\sqrt{5})/{10}$ at the points $(1-\sqrt{5})/{2}$ and $(1+\sqrt{5})/{2}$. Note that this distribution satisfies $E(W_i) = 0$, $E(W^2_i) = E(W^3_1)=1$. 

\noindent \textbf{Step 4.} Let $D^{\ast b}$ be the test statistic obtained from the bootstrap samples ${ \left( {X_{ij}, Y_{ij}^{\ast b}} \right)}_{i=1}^{n_j}$, $i = 1, \ldots, n_j$, $j = 1, \ldots, J$  after applying the steps 1 and 2 to the cited bootstrap sample. 

As we mentioned, the decision rule consists of rejecting the null hypothesis if $D > \widehat D^{(1-\alpha)}$, where $\widehat D^{(1-\alpha)}$ is the empirical $(1-\alpha)$-percentile of  the values $D^{\ast 1}, \ldots, D^{\ast B}$ obtained before.

It is important to note that the described procedure -- testing $H_0(K)$-- should be repeated from $K = 1$ onwards until a certain null hypothesis is not rejected in order to determine  automatically the number  of $K$ groups. Note, however, that unlike the previous test decision, this latter one is not statistically significant (strong evidences for rejecting the null hypothesis are not given). The whole procedure is briefly described step by step in Algorithm \ref{algoreg}. 

\begin{algorithm}
\caption{$K$-regression curves  algorithm}
\label{algoreg}
\begin{enumerate}
\item  With ${ \left( {X_{ij}, Y_{ij}} \right)}_{i=1}^{n_j}, i=1, \ldots, n_j, j = 1, \ldots, J$,  obtain $\hat m_j$.
\item Initialize with $K = 1$ and test $H_0(K)$:
\begin{enumerate}
\item[2.1] Obtain the ``best'' partition $G_1, \ldots, G_K$ by means of the $K$-means or $K$-medians algorithm.

\item[2.2] For $k = 1, \ldots, K$, estimate $\widehat{f}_k$ and retrieve the test statistic $D$.
\item[2.3] Generate $B$ bootstrap samples and calculate $D^{\ast b}$, for $b = 1, \ldots, B$.
\item[2.4] \textbf{if} $D > D^{\ast (1-\alpha)}$ \textbf{then} 
	\subitem reject $H_0(K)$
	\subitem $K = K + 1$
	\subitem go back to 2.1 \\
	\textbf{else}
	\subitem accept $H_0(K)$\\
	\textbf{end}
\end{enumerate}	
\item The number $K$ of groups of  equal regression curves is determined.
\end{enumerate}
\end{algorithm}


\section{Simulation study}
\label{noramv:simulationest}

Results of three Monte Carlo experiments settings conducted to evaluate the finite sample performance of the proposed methodology are reported in this section.  Firstly, we show  those ones related with testing one specific hypothesis $H_0(k)$, particularly, we start with $K = 1$. Note that testing $H_0(1)$ is equivalent to test the null hypothesis of no difference in nonparametric regression between two or more independent groups. Accordingly, we will compare our procedure with some other methods described in literature to this end, particularly with those ones used in  \cite{Pardo} and \cite{park2014}.    

Based on these two publications, the following scenarios (with models and variance functions) are proposed in order to carry out the simulation:

\bigskip

\noindent (R1) $m_1(x) = m_2(x) =m_3(x) = x $

\noindent  (R2)  $m_1(x) = x, m_2(x) = x + 0.25, m_3(x) = x + 0.5 $

\noindent (R3)  $m_1(x) =x,  m_2(x) = 0.5, m_3(x) = 1 -x$

\noindent (R4)  $m_1(x) = m_2(x) =x, m_3(x) = 1 - 48 x + 218 x^2 - 315 x^3 + 145 x^4$

\medskip

\noindent (V1)  $\sigma^2_1(x) = \sigma^2_2(x) =\sigma^2_3(x) = 0.5$

\noindent (V2)  $\sigma^2_1(x) = \sigma^2_2(x) =  \sigma^2_3(x) =  0.5(0.5 + 2x) $

\noindent (V3)  $\sigma^2_1(x) =x,  \sigma^2_2(x) = 0.5, \sigma^2_3(x) = 0.5(2.5 - 2x)$

\noindent (V4)  $\sigma^2_1(x) = \sigma^2_2(x) =x, \sigma^2_3(x) = 0.5(-4x^2 + 4x+ 0.5).$

\bigskip

\noindent Note that, all the above scenarios were considered in the simulation section described by \cite{park2014}  while only (R1), (R2), (R3) models with variance function (V1) were considered in \cite{Pardo}. 

The simulated data were generated from equation in (\ref{noramv:eq1}), with 
the covariate $X_j$  drawn from a uniform distribution on the interval $[0,1]$ and with independent model errors $\varepsilon_{j}$   drawn from a standard normal distribution with mean 0 and variance $\sigma^2_j(x)$, for $j = 1,2, 3$. In each case, to determine the critical values of the test statistics we applied bootstrap method, specifically using $B = 500$ bootstrap samples. In order to perform an unbalanced study we used unequal sample sizes for each $j$ curve, with $n_j = 300, 400, 500$ . Both type I error rates and power values were calculated on the basis of 1000 simulation runs at the significance levels of $\alpha =0.05$ and $\alpha = 0.10$.  Finally, note that bandwidths are selected automatically by minimising the cross-validation criterion defined in Section \ref{noramv:methodology}.

Table \ref{noramv:PardoPark} shows the results under the null hypothesis  --(R1) model-- and under the alternative --(R2), (R3), and (R4) models-- of the tests based on $D_{CM}$ and $D_{KS}$. Note that errors are homoscedastic  when variance function (V1) is chosen and heteroscedastic when we select the variance functions (V2)-(V4).

 The two test statistics of our procedure control type I error rate very close to the nominal level and this approximation is much better than  the  obtained by \cite{park2014} with their SiZer method. This fact is achieved for all variance functions. Unlike our method, the  SiZer makes less mistakes than it should be with mean type I errors  below 0.004 quite far from $\alpha = 0.05$. Focussing on the comparison of our two proposed test statistics, $D_{CM}$ seems to obtain better approximations than $D_{KS}$.

The results in terms of power performance for the alternatives are good reaching the value of 1 for all cases. These power values are higher than those obtained by \cite{park2014}  whose values range between 0.5090 to 0.7312. For all cases, the four types of variance functions do not represent a big effect on the rejection probabilities.


 \begin{table}[h]
\centering
\caption{Proportion of rejection probabilities for the null hypothesis $H_0(1)$ of the tests based on $D_{CM}$ and $D_{KS}$. The models are homoscedastic, with variances given in (V1), and heteroscedastic, with variances given from (V2) to (V4). The significance level is $\alpha = 0.05, 0.10$ and the sample size is $n_j = 300, 400, 500$.}
\begin{tabular}{llrrrrccrrc}
  \hline
 & \multicolumn{4}{c}{$D_{CM}$} &   & \multicolumn{4}{c}{$D_{KS}$}      \\
\cline{3-6} \cline{8-11} 
 $\alpha$ &  & (V1) & (V2) & (V3) & (V4) & & (V1) & (V2) & (V3) & (V4)   \\ 
  \hline
\multirow{1}{*}{0.05}  & (R1) & 0.041 & 0.049 & 0.052 & 0.052 &  & 0.056 &  0.043& 0.050 & 0.054   \\ 
   &  (R2) & 1.000& 1.000 & 1.000 & 1.000 && 1.000& 1.000 & 1.000 & 1.000  \\ 
   &  (R3)& 1.000 & 1.000 & 1.000 & 1.000  && 1.000 & 1.000 & 1.000 & 1.000  \\ 
   & (R4) &  1.000& 1.000 & 1.000& 1.000 && 1.000 &  1.000&  1.000&  1.000 \\ 
   \hline
\multirow{1}{*}{0.10}  & (R1) &0.088 & 0.088 & 0.091 & 0.093  && 0.086&0.088 &  0.107&  0.099   \\ 
   &  (R2) & 1.000& 1.000 & 1.000 & 1.000 && 1.000 & 1.000 & 1.000 & 1.000  \\ 
   &  (R3)& 1.000& 1.000 & 1.000 & 1.000  && 1.000 & 1.000 & 1.000 &  1.000 \\ 
   & (R4) & 1.000& 1.000 & 1.000 & 1.000 && 1.000 & 1.000 & 1.000 & 1.000  \\ 
   \hline
\end{tabular}
\label{noramv:PardoPark}    
\end{table}

When comparing the results reported in Table \ref{noramv:PardoPark}  with those ones obtained by  \cite{Pardo} --from (R1) to (R3)-- very similar behaviours  both in type I errors and powers were found. If minor differences  are appreciated in terms of power these could be due to  sample size. The highest sample size used by these authors was  $n_j = 100$ against size of $n_j = 500$ used in this simulation.

The second simulation setting was designed to assess the performance of the procedure testing one specific hypothesis $H_0(K)$, using in this case $K =5$. Note that, in this case, we are dealing  with testing if the $J$ regression curves can be grouped in five groups.  
Accordingly, a new scenario is proposed. The regression model given in (\ref{noramv:eq1}) is considered for $j=1, \ldots, 30$, with

\[m_j(X_j)=\left \{ {
\begin {tabular} {lcc}
$X_j + 2 $ \hspace{1.85cm} {\text if} \quad $j \le 5 $\\
$X_j^2 + 3$ \hspace{1.80cm} {\text if} \quad $5 < j \le 10 $\\
$2 \hspace{0.10cm} {\rm sin} \hspace{0.05cm}(2 \hspace{0.10cm}X_j) - 2$ \hspace{0.5cm} {\text if} \quad $10 < j \le 15 $\\
$2 \hspace{0.10cm} {\rm sin}\hspace{0.05cm}(X_j)$ \hspace{1.4cm} {\text if} \quad $15 < j \le 20 $\\
$2 \hspace{0.10cm} {\rm sin} \hspace{0.05cm}(X_j) + a \hspace{0.10cm} \mathrm{e}^{X_j}$ \hspace{0.08cm} {\text if} \quad $20 < j \le 25 $\\
$1$  \hspace{2.65cm} {\text if} \quad $j > 25 $,\\
\end {tabular}
}\right.\]
\noindent
where $a$ is a real constant, $X_j$ is the explanatory covariate drawn from a   uniform distribution on the interval$[-2,2]$, and  $\varepsilon_j$ is the error distributed in accordance to a $N(0, \sigma^2_j(x))$. We have also considered a homoscedastic and a heteroscedastic situation. In the homoscedastic case, the variance functions are given by $\sigma^2_j(x) = 0.5$, while in the heteroscedastic case, the variance functions are given by $\sigma^2_j(x)=0.5+0.05m_j(x)$.

As in the previous simulated scenario, $B=500$ bootstrap samples were generated in order to know the distribution of the test statistic.  Type I error rates and power values are calculated as the proportions of rejections in 1000 simulation  for different significance levels ($\alpha = 0.05, 0.10$). We have  also considered  unequal sample sizes for each $j$ curve, particularly, $(n_1, n_2, \ldots, n_J) \sim$ Multinomial$(n;  p_1, p_2, \ldots, p_J)$  being $p_j = p_j^\ast /\sum_{j=1}^{J} p_j^\ast$, with  $p_j^\ast$ randomly drawn from $\{1, 1.5, 2, 2.5, 3\}$. We  used  $n = 1000, 3000$ and $6000$. 

Different values of $a$ were considered, ranging from 0 to 0.4. It should be noted that the value  $a=0$ corresponds to the null hypothesis (the thirty regression curves can be classified in five groups), and as the value of $a$ increases, so does the difference between the curves leading to six groups.

\begin{figure} 
  \begin{center} 
\includegraphics[width=\textwidth]{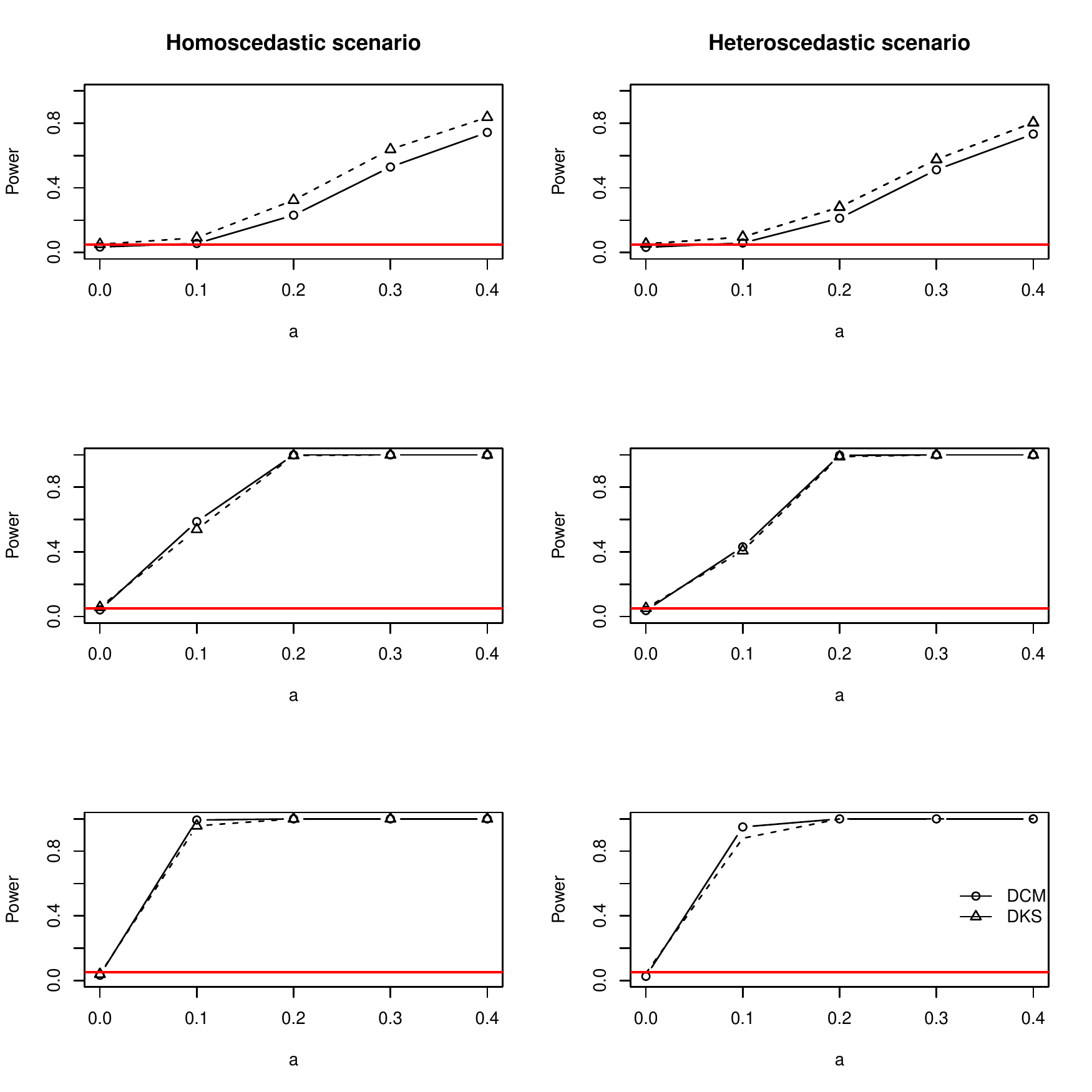}
    \caption{Power curves. Rejection probabilities of the proposed tests  for nominal level 5\% (red line), for $n=$ 1000, 3000 and 6000 (upper, middle and lower panels, respectively) and for both scenarios (left and right panels, respectively)}
    \label{noramv:figura} 
  \end{center} 
\end{figure}

Type I errors are  registered by using the test statistics $D_{CM}$ and $D_{KS}$ for different significance levels and sample sizes in Table \ref{noramv:homo_hetero}. Results reported in this Table \ref{noramv:homo_hetero} reveal that the the two test statistics perform similarly and reasonably well, with the level being held or coming fairly close to the nominal size in most cases, especially with large sample sizes. Some test performance results in terms of power are shown in Table \ref{noramv:powerhomohetero} and Figure \ref{noramv:figura}. Both in the homoscedastic  and in the heteroscedastic  situation, the behaviour of the power is good, although perhaps it seems somewhat better in the homoscedastic situation. As expected, the power  improves as the sample size grows. The highest values of power  were achieved with the Cr\'{a}mer Von Misses type test based on the $L_2$-norm.

\begin{table}
\begin{center}
 \caption{Estimated type I error of testing $H_0(5)$ based on the test statistics $D_{CM}$ and $D_{KS}$, for different sample sizes and nominal levels. Results given for the homoscedastic and the heteroscedastic situation.}
\label{noramv:homo_hetero}  
\begin{small}
\begin{tabular}{ccccccccccc }
\noalign{\smallskip}
\hline\noalign{\smallskip}
&&\multicolumn{2}{c}{$n$ = 1000}&&\multicolumn{2}{c}{$n$ = 3000}&&\multicolumn{2}{c}{$n$ = 6000}&  \\
\cline{3-4}\cline{6-7} \cline{9-10}\noalign{\smallskip}
Scenario&$\alpha$:&0.050 & 0.100 &  &0.050  & 0.100 && 0.050 & 0.100 &        \\
\cline{1-4}\cline{6-7} \cline{9-10}\noalign{\medskip}
\multirow{2}{*}{Homoscedastic} &$D_{CM}$& 0.034 & 0.067 &  &  0.041 & 0.080  & &0.033 & 0.071 &         \\ 
   & $D_{KS}$& 0.051 & 0.104  &  &  0.057 & 0.104  & &  0.039 & 0.088 &      \\ 
\noalign{\medskip}
  \hline
\noalign{\medskip}
\multirow{2}{*}{Heteroscedastic} & $D_{CM}$& 0.032 & 0.083 &  &0.037 & 0.079 && 0.026 & 0.076 &\\ 
             & $D_{KS}$ & 0.053 & 0.113 &  & 0.051& 0.109 & & 0.045  & 0.108 &        \\ 
\noalign{\medskip}
\hline\noalign{\smallskip}
\end{tabular}
\end{small}
\end{center}
\end{table}

\begin{table}
\begin{center}
\caption {Rejections probabilities of testing $H_0(5)$ based on the test statistic   $D_{CM}$  and  $D_{KS}$ for different $a$ values and different sample sizes. The significance level is $\alpha = 0.05,0.10$. Results given for the homoscedastic and heteroscedastic scenario.}
\label{noramv:powerhomohetero}
\begin{tabular}{rrclrrcrrr}
 \hline\noalign{\smallskip}
&   & &  \multicolumn{3}{r}{Homoscedastic} & & \multicolumn{3}{r}{Heteroscedastic}\\
\cline{4-6}\cline{8-10}\noalign{\medskip}
$n$ & a  & &$\alpha$:& 0.050 & 0.100& &0.050 & 0.010\\
\hline

\multirow{8}{*}{1000}  & \multirow{2}{*}{0.1} & $D_{CM}$ & & 0.056 & 0.129 && 0.059 & 0.135 \\
 &  &  $D_{KS}$ &&   0.091 & 0.191 && 0.096 & 0.177\\
  \cline{2-10}
  & \multirow{2}{*}{0.2}   & $D_{CM}$ & &0.231 & 0.359  &&0.212 & 0.344\\
&  &  $D_{KS}$ && 0.324 & 0.457 && 0.281 & 0.402\\
  \cline{2-10}\noalign{}
  & \multirow{2}{*}{0.3}   & $D_{CM}$ & & 0.529 & 0.650  &&0.512 & 0.639 \\
&  &  $D_{KS}$ && 0.638 & 0.746 && 0.576 & 0.696\\
\cline{2-10}\noalign{}
  & \multirow{2}{*}{0.4}   & $D_{CM}$ & &0.743 & 0.813  &&0.733 & 0.816\\
&  &  $D_{KS}$ &&  0.837 & 0.887 &&0.804 & 0.862\\
\hline\noalign{\smallskip}
\multirow{8}{*}{3000}  & \multirow{2}{*}{0.1} &$D_{CM}$ & & 0.587 & 0.719 && 0.431 & 0.607 \\
 &  &  $D_{KS}$ &&   0.539 & 0.682  &&0.407 & 0.584 \\
  \cline{2-10}
  & \multirow{2}{*}{0.2}   & $D_{CM}$ & & 1.000 & 1.000 &&0.997 & 0.999\\
&  &  $D_{KS}$ && 0.996 & 0.999 &&0.989 & 0.997 \\
  \cline{2-10}\noalign{}
  & \multirow{2}{*}{0.3}   & $D_{CM}$ & & 1.000 & 1.000 &&1.000 & 1.000\\
&  &  $D_{KS}$ && 1.000 & 1.000 &&1.000 & 1.000\\
\cline{2-10}\noalign{}
  & \multirow{2}{*}{0.4}   & $D_{CM}$ & & 1.000 & 1.000 &&1.000 & 1.000\\
&  &  $D_{KS}$ && 1.000 & 1.000 &&1.000 & 1.000\\
\hline\noalign{}
\multirow{8}{*}{6000}  & \multirow{2}{*}{0.1}   & $D_{CM}$ & & 0.993 & 0.996  && 0.950 & 0.980\\
 &  &  $D_{KS}$ &&  0.957 & 0.980  && 0.880 & 0.927\\
  \cline{2-10}
  & \multirow{2}{*}{0.2}   & $D_{CM}$ & &1.000 & 1.000  &&1.000 & 1.000 \\
&  &  $D_{KS}$ && 1.000 & 1.000 &&1.000 & 1.000 \\
  \cline{2-10}\noalign{}
  & \multirow{2}{*}{0.3}   & $D_{CM}$ & & 1.000 & 1.000   &&1.000 & 1.000 \\
&  &  $D_{KS}$ && 1.000 & 1.000 &&1.000 & 1.000 \\
\cline{2-10}\noalign{}
  & \multirow{2}{*}{0.4}   & $D_{CM}$ & &1.000 & 1.000  &&1.000 & 1.000 \\
&  &  $D_{KS}$ && 1.000 & 1.000 &&1.000 & 1.000 \\
\hline\noalign{}
\end{tabular}
\end{center}
\end{table}

Finally, note that in order to obtain the results of the  simulations, we have used wild bootstrap and we have  selected the bandwidth (both in the original samples and in the bootstrap replicates) by means of cross-validation. However, although the whole results are not shown, we have also evaluated both the selected bootstrap procedure  and the effect of the smoothing parameter. Regarding to the resampling technique, the simple bootstrap was also used; and  two new situations were considered for the bandwidth: (i) to use  cross-validation for obtaining the bandwidth for the original samples and then to fix these in the bootstrap replicates, (ii)  to fix some bandwidths a priori. In all cases,  poor results were obtained (lower powers and poorer approximations of type I error, especially with small sample sizes) when compared to those based on the the approximation exposed here.

The last conducted simulation aims to assess the performance of the Algorithm \ref{algoreg}, i.e., the whole procedure. We compare our method with the procedure recently developed by \cite{vogt2016}. The method aims to classify nonparametric functions in the longitudinal data framework.  We have kept exactly the scenario described in Section 5 of the above mentioned paper. Then,  for $j=1, \ldots, 120$, the regression model given in (\ref{noramv:eq1})  is considered with

\[m_j(X_j)=\left \{ {
\begin {tabular} {lcc}
$0 $ \hspace{4.85cm}  {\text if}\quad $j \le 50$,\\
$1 - 2X_j$ \hspace{3.80cm} {\text if} \quad $51 < j \le 80 $,\\
$0.75 \hspace{0.10cm} {\rm tan^{-1}} \hspace{0.05cm}10 (\hspace{0.10cm}X_j - 0.6)$ \hspace{1.3cm} {\text if} \quad $81 < j \le 100$,\\
$2.5  \hspace{0.10cm} {\rm}  (1 - X_j^2)^4 \hspace{0.05cm}\mathds{1} (|{X_j}| \le 1)$ \hspace{1.13cm} {\text if} \quad $101 < j \le 110$,\\
$1.75 \hspace{0.10cm} {\rm tan^{-1}} \hspace{0.05cm}5 (X_j - 0.6) + 0.75 $ \hspace{0.53cm} {\text if} \quad $111 < j \le 120$,\\
\end {tabular}
}\right.\]
\noindent
being  $X_j$  the explanatory covariate drawn from a  uniform distribution  on the interval $[0,1]$, and  $\varepsilon_j$  the error distributed in accordance to a normal distribution $N(0, 1.3)$. 
The  simulation study was carried out under different sample sizes $n_j = 100, 150, 200$ taking into account the test statistic $D_{CM}$. The remainder parameters of the simulation (number of simulation runs and number of  bootstrap replicates) were kept as in the previous two scenarios.

In order to perform correctly, the Algorithm \ref{algoreg} must reject the first null hypothesis, $H_0(1)$, continues, rejects again the second one, $H_0(2)$, and so on  until it  accepts $H_0(5)$.   Results of this simulation are shown in Table \ref{noramv:vogt16} which refers to  the number of times that the procedure works well (in \%) selecting the number of groups $K$ and using a nominal level of 5 \%. 


\begin{table}
\centering
\caption{Number of times  in \% (of 1000 repetitions) that the Algorithm \ref{algoreg}  selects the number of groups using a nominal level of 5\%.}  
\begin{tabular}{lrrrrrrrrrrrrrrr}
  \hline
& & \multicolumn{6}{c}{Number of groups}          \\
\cline{2-8}  
$n_j$ & 4 & 5 & 6 & 7 & 8 & 9 & 10  \\ 
  \hline
 \multirow{1}{*}{100}  &   0.2 & 94.3 & 4.1 & 1.1 & 0.1 & 0.1 & 0.1  \\ 
      \hline
  \multirow{1}{*}{150}  &  0.0 & 94.5 & 4.4 & 0.9 & 0.1 & 0.1 & 0.0  \\ 
      \hline
\multirow{1}{*}{200}  &  0.0 & 95.2 & 3.7 & 0.7 & 0.4 & 0.0 & 0.0  \\ 
      \hline
\end{tabular}
\label{noramv:vogt16}    
\end{table}

 
  Results reported in Table \ref{noramv:vogt16} reveal a good behaviour of the proposed Algorithm 1, with rates of success  around 95\%, coming quite close to the $(1 - \alpha)$ established. Already for the smallest samples size $n_j = 100$, our procedure selects the true number of groups $K=5$ in 94.3\% of the times unlike the results obtained by the method proposed in \cite{vogt2016} that show the 75\% of the times. Moreover, there is a slight improvement on the proportion of success as the sample size increases.

\begin{figure}
  \begin{center} 
 \includegraphics[width=\textwidth]{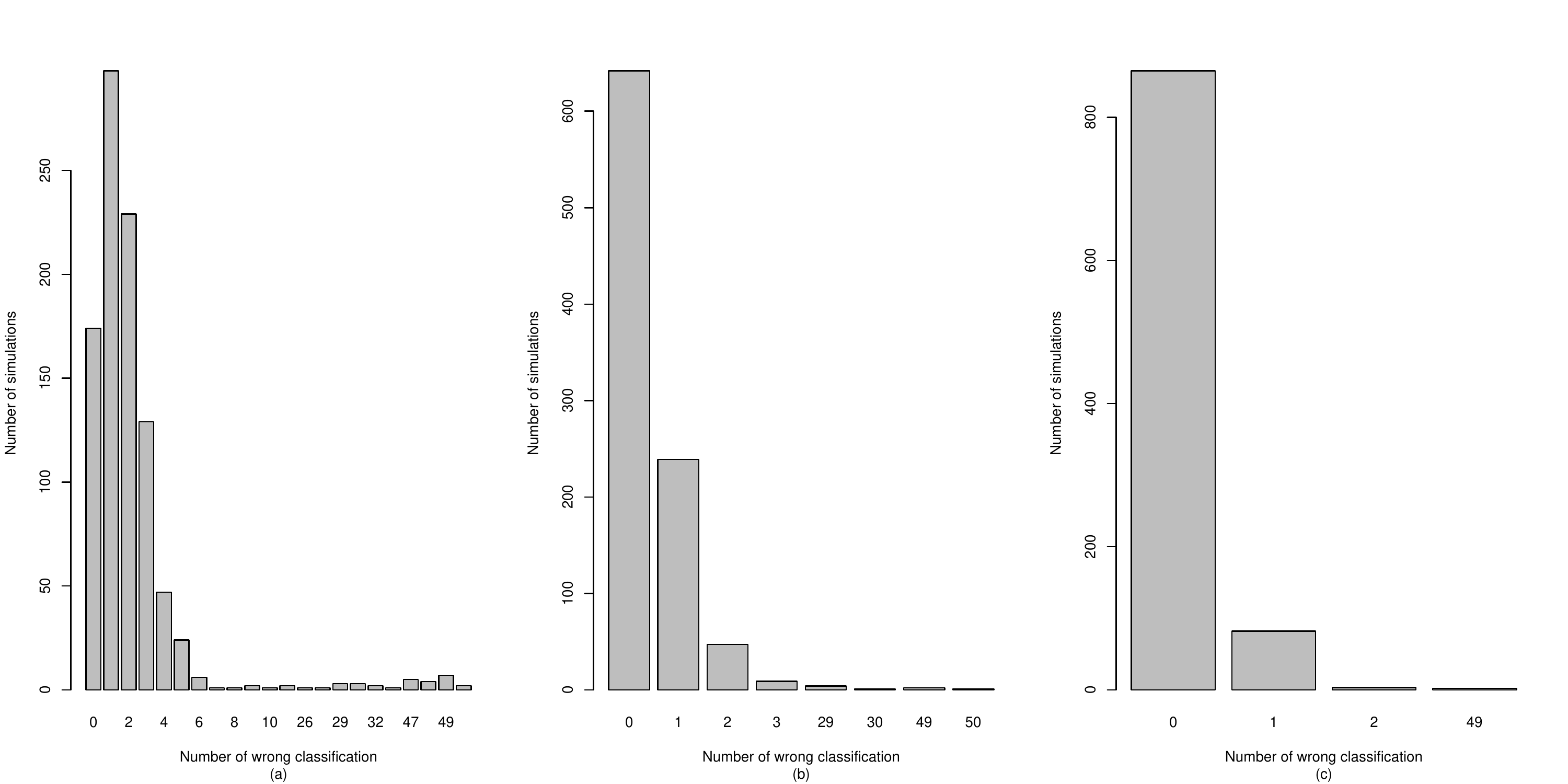} 
   \caption{Simulation results for the estimation of the groups $G_1, \dots, G_5$.  (a)--(c) distributions of the number of wrong assignments for the Algorithm \ref{algoreg}  (without Holm's correction) and sample sizes $n_j = 100, 150, 200$, respectively}  
    \label{noramv:figwrong} 
  \end{center} 
\end{figure}

To measure how well the  procedure assigns each curve ($J=120$)  to their correct group ($G_1, \ldots, G_5$), the number of curves wrongly classified was analysed. Figure \ref{noramv:figwrong} shows the distribution of this variable for the different sample size ($n_j = 100, 150, 200$).  Particularly, it is shown the number of times  in which a certain number of wrong assignments is obtained. Note that as the sample size increases, so does the number of correctly classified curves. For  $n_j = 100$  our procedure gives satisfactory results taking into account that  at most five curves are  wrongly classified  in 90\% of cases.  At a sample size of $n_j = 150$ our procedure is able to  classify correctly all the curves in about 70\% of the cases. Finally, for the biggest sample size, all the curves are correctly classified  in 91\% of the cases and most of the wrong classifications are due to the error in only one single curve out of a total of 120.  It should be mentioned that these results are quite better regarding to those ones obtained by \cite{vogt2016} in which, considering  the best scenario ($n_j = 200$), all the curves are correctly classified about 80\% of cases.

\section{Application to real data}
\label{noramv:app}

Our  methodology is applied to the study of the geometry of a tunnel through the analysis of a set of cross-sections. These sections  were obtained by adjusting a point cloud collected with a  RIEGL LMS.Z390 time of flight  terrestrial laser scanner (TLS). It has a maximum range of about 400 m for objects with reflectance greater than $80\%$ and 6 mm nominal accuracy for a single point to a distance of 50 m. The capture rate is 8000 data points/second.

Terrestrial laser scanning is a ground based technique to measure the position and dimension of objects in a three dimensional space. Thereby, a laser beam is emitted from a laser light source and used to scan the surface of surrounding objects. The distance between the scanner and the object is determined by the time of flight principle. The laser range finder sends a laser pulse towards the object and measures the time taken by the pulse to be reflected from the target and returned to the sender. Also horizontal and vertical angles from the center of the scanner to the point are measured. Then the coordinates of the points are calculated using polar coordinates.

The tunnel was made by drilling and blasting. It is approximately circular and its theoretical diameter  is  9 m. The cross-section of the tunnel should be constant throughout the tunnel but, in drill and blast, there can be over-break amounting to 10 to $20\%$ of the excavated cross-sectional area, which must be removed and possibly refilled. In the case of an insufficient excavation, the cross-section must be expanded by mechanical or manual methods. This represents a significant increase in the cost of the work.

\begin{figure}
  \begin{center} 
  \includegraphics[width=\textwidth]{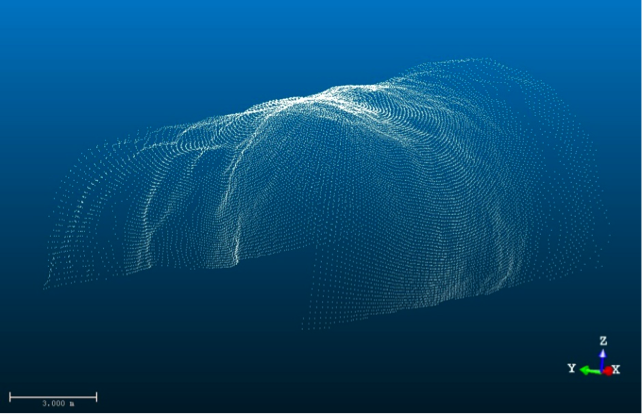}
     \caption{Tunnel point cloud. Tunnel  measured with the terrestrial laser scanner} 
    \label{noramv:fig4} 
  \end{center} 
\end{figure}

Figure \ref{noramv:fig4} shows the point cloud of the tunnel obtained with the TLS. Although the irregularities of its surface can be visually appreciated, it is of interest to know if the tunnel section is, in general terms, homogeneous or if, on the contrary, there are different areas of homogeneous section. The engineers could take advantage of this information to plan activities to rebuild the tunnel. Thus, it is very interesting   to distinguish between areas of the tunnel that need to be refilled from those that are under-excavated and, consequently, need to be widened. It also can provide useful information regarding the mechanical characteristics of the materials. Moreover, our methodology allows taking into account the fact that some differences between the profiles could be due to the noise associated to the point cloud. Then small differences between the profiles that could have their origin in the measuring system are not computed as real differences in the tunnel.

In order to determine the homogeneity of the tunnel, a data set  of $n=$ 16 075 coordinates constituted by  66 cross-sections were obtained from the point cloud in Cartesian coordinates. The cross-sections were obtained by the intersection of the plane perpendicular to the axis of the tunnel and the 3D model created from the point cloud. For $j= 1, \ldots, 66$  sections, the point cloud measured with the TLS provides $(X_j,Y_j)$ coordinates. As the cross-sections given in Cartesian coordinates are not univariate functions, we first perform a  transformation of the coordinates. The Cartesian coordinates $(X_j,Y_j)$ can be converted  to polar coordinates $(r_j,\alpha_j)$ given by 

\[
(r_j,\alpha_j) =\left({ \sqrt{X_j^2 + Y_j^2)}, 
\arctan  \left ({ {Y_j}/{X_j} }\right)}\right)
\]

\noindent being $r$ the distance to each point and $\alpha$ the polar angle.  Then, we  use our proposed  approach in the following regression models

\[r_j=m_j(\alpha_j) + \varepsilon_j \quad \text{for} \quad j=1, \ldots, 66.\]

The application of the proposed methodology ---using the test statistics $D_{CM}$--- indicates that, for a nominal level of 5\%, the null hypothesis is rejected (with p-values lower than 0.01) until $K=5$ (with a p-value of 0.45) and therefore the cross-section $m_1, \ldots, m_{66}$ can be grouped in five groups.

\begin{figure}
  \begin{center} 
 \includegraphics[width=\textwidth]{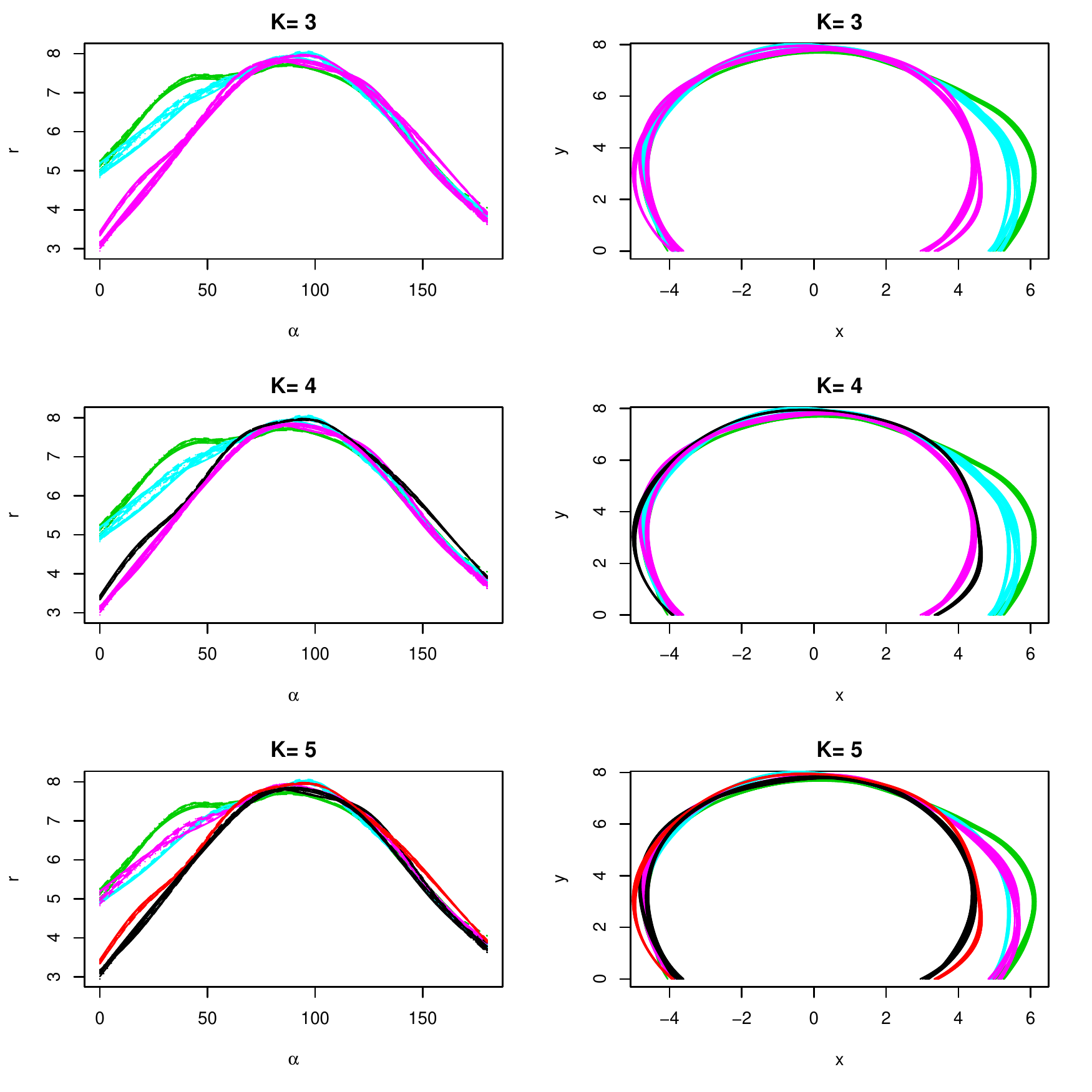} 
    \caption{Estimated tunnel regression curves. The curves are colored according to the groups to which they belong. Upper, middle and lower panels correspond  to these curves assigned into three ($K=3$), four ($K=4$) and five groups ($K=5$), respectively. Left and right  panels are the  estimates in polar and Cartesian coordinates, respectively.}  
    \label{noramv:fig2} 
  \end{center} 
\end{figure}

The estimated regression curves in polar coordinates given by $(\alpha_j,\hat m_j(\alpha_j))$ for $j=1,\ldots,66$ are drawn in the left panel of Figure \ref{noramv:fig2}. Curves assigned to each group are plotted with  the same color. The differences between  groups are almost visually appreciable. Additionally, in order to graphically assess the  analysis, the curves are shown  in Cartesian coordinates using the transformation $\left({\hat m_j(\alpha_j) \cos (\alpha_j),\hat m_j(\alpha_j) \sin (\alpha_j) }\right)$  (see right panel of  Figure \ref{noramv:fig2}).

\begin{figure}
  \begin{center} 
\includegraphics[width=\textwidth]{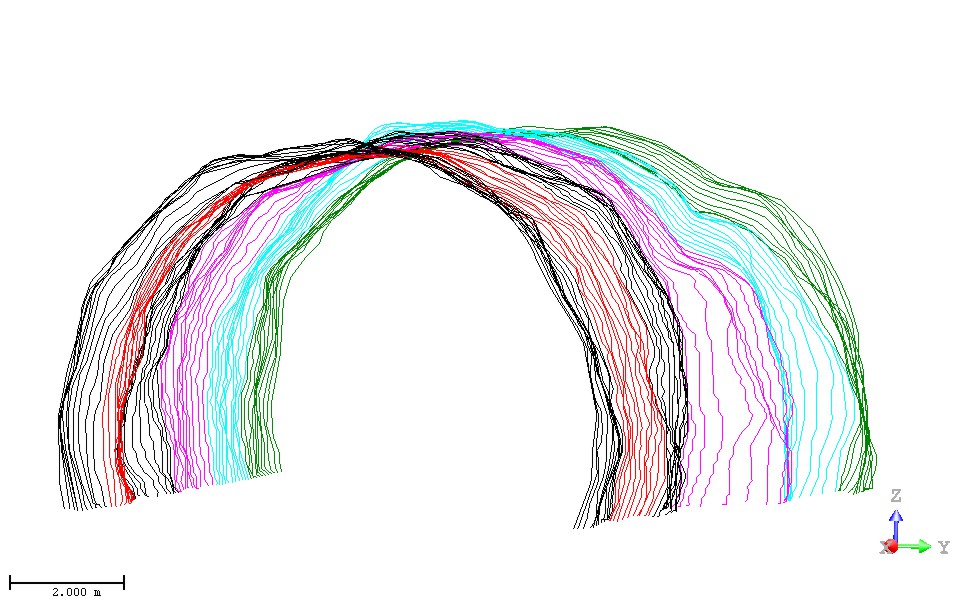}
   \caption{Spatial distribution of  tunnel profiles. A specific color is assigned for each section according to the group to which it belongs (in this case five groups, $K=5$).} 
    \label{noramv:fig5} 
  \end{center} 
\end{figure}

Finally, Figure \ref{noramv:fig5} depicts the spatial distribution of the cross-sections along the tunnel. A specific color is assigned for each section according to the group in which it belongs.  As can be appreciated, there are two non-consecutive areas that belong to the same group (printed in black). It is also clear that three of the five groups (magenta, blue and green profiles) correspond to areas of the tunnel with a big over-excavation. The differences between them tell us that different coatings of shotcrete are needed for each of the three~areas.

The good performance of  the proposed procedure in terms of spatial distribution should be highlighted. The cross-sections which are spatially close   belong to the  same group, and accordingly, this provides  an easy  interpretation of results.   Therefore, the identification of homogeneous areas allows a better planning of the work to build the final section of the tunnel, which should fit with the theoretical one.

\section{Conclusion}
\label{noramv:conclusion}

Several nonparametric procedures have been proposed for the comparison of regression functions, however they are no fully informative when the null hypothesis is rejected. This is especially important in practice and in situations where the number of  $J$ curves is large enough. In order to solve this situation, we proposed a new procedure that let us not only testing the equality of nonparametric regression curves but also grouping them if they are not equal. Insofar as our comprehensive simulation studies are concerned, we have shown that the results  are satisfactory (both in terms of  type I error and power) and indicate similar or even better performance than other  procedures used in the literature.

The methodology has been applied to a tunnel inspection work addressed to determine heterogeneities in the tunnel geometry. By grouping a set of cross-sections of the tunnel surface it has been possible to identify five areas with significant differences in their geometry. 

It is worth mentioning that software in the form of an R package named clustcurv \cite{nora_clustcurv,clustcurv} was developed implementing the proposed method which seems to be  stable and computational efficient. clustcurv is available from the Comprehensive R Archive Network  (CRAN) at \url{https://cran.r-project.org/web/packages/clustcurv} and can be called to group multiple nonparametric curves both in the regression and  in the survival framework.

Although the proposed test is designed for detecting groups of regression curves with continuous response, it seems that the extension to other responses such as Bernoulli, Poisson or other parametric families can be achieved without much effort. Moreover, our methodology suffers from the limitation of solely being able to address a single continuous covariate, few further work is nevertheless needed to extend the proposed methodology to the case of multiple covariates.

Even though two different clustering techniques have been implemented in the algorithms proposed to determine groups of curves, the reach of this work might be expanded allowing more clustering techniques such as K-medoids \citep{books/wi/KaufmanR90} or Mean-Shift algorithms \citep{meanShift}. Finally, a more challenging target is to consider the application of these methods to the Big Data environment where the classical bootstrap techniques can be prohibitively demanding.

%
%
%
%
%
%
%
%

\bibliographystyle{apa}
\bibliography{bibliografia}  

\begin{thebibliography}{}

\bibitem[\protect\astroncite{Abraham et~al.}{2003}]{abraham03}
Abraham, C., Cornillon, P.~A., Matzner-L{\o}ber, E., and Molinari, N. (2003).
\newblock Unsupervised curve clustering using b-splines.
\newblock {\em Scandinavian Journal of Statistics}, 30(3):581--595.

\bibitem[\protect\astroncite{de~Boor}{2001}]{deboor}
de~Boor, C.~A. (2001).
\newblock {\em A Practical Guide to Splines}.
\newblock Springer Verlag, New York.

\bibitem[\protect\astroncite{Delgado}{1993}]{Delgado}
Delgado, M.~A. (1993).
\newblock Testing the equality of nonparametric regression curves.
\newblock {\em Statistics and Probability Letters}, 17:199--204.

\bibitem[\protect\astroncite{Dette and Neumeyer}{2001}]{dette2001}
Dette, D. and Neumeyer, N. (2001).
\newblock Nonparametric analysis of covariance.
\newblock {\em The Annals of Statistics}, 29:1361--1400.

\bibitem[\protect\astroncite{Efron}{1979}]{NoRefWorks:8}
Efron, B. (1979).
\newblock {Bootstrap methods: another look at the jackknife}.
\newblock {\em The Annals of Statistics}, 7:1--26.

\bibitem[\protect\astroncite{Efron and Tibshirani}{1993}]{NoRefWorks:9}
Efron, E. and Tibshirani, R.~J. (1993).
\newblock {\em {An introduction to the Bootstrap}}.
\newblock Chapman and Hall, London.

\bibitem[\protect\astroncite{Fan and Gijbels}{1996}]{GVK19282144X}
Fan, J. and Gijbels, I. (1996).
\newblock {\em Local polynomial modelling and its applications}.
\newblock Number~66 in Monographs on statistics and applied probability series.
  Chapman {\&} Hall.

\bibitem[\protect\astroncite{Fan and Marron}{1994}]{NoRefWorks:12}
Fan, J. and Marron, J. (1994).
\newblock {Fast implementation of nonparametric curve estimators}.
\newblock {\em Journal of Computational and Graphical Statistics}, 3:35--56.

\bibitem[\protect\astroncite{Garc{\'\i}a-Escudero and
  Gordaliza}{2005}]{garciaescudero05}
Garc{\'\i}a-Escudero, L.~A. and Gordaliza, A. (2005).
\newblock A proposal for robust curve clustering.
\newblock {\em Journal of Classification}, 22(2):185--201.

\bibitem[\protect\astroncite{Golub et~al.}{1979}]{golub}
Golub, G., Heath, M., and Wahba, G. (1979).
\newblock Generalized cross-validation as a method for choosing a good ridge
  parameter.
\newblock {\em Technometrics}, 21(2):215--223.

\bibitem[\protect\astroncite{Gonz{\'a}lez-Manteiga and Cao}{1993}]{cao}
Gonz{\'a}lez-Manteiga, W. and Cao, R. (1993).
\newblock Testing the hypothesis of a general linear model using nonparametric
  regression estimation.
\newblock {\em Test}, 2(1):223--249.

\bibitem[\protect\astroncite{Gonz{\'a}lez-Manteiga and
  Crujeiras}{2013}]{crujeiras2013}
Gonz{\'a}lez-Manteiga, W. and Crujeiras, R.~M. (2013).
\newblock {An updated review of Goodness-of-Fit tests for regression models}.
\newblock {\em Test}, 22:361--411.

\bibitem[\protect\astroncite{Hall and Hart}{1990}]{hall1990}
Hall, P. and Hart, J.~D. (1990).
\newblock Bootstrap test for difference between means in nonparametric
  regression.
\newblock {\em Journal of the American Statistical Association},
  85(412):1039--1049.

\bibitem[\protect\astroncite{H{\"a}rdle and Mammen}{1993a}]{NoRefWorks:18}
H{\"a}rdle, W. and Mammen, E. (1993a).
\newblock Comparing nonparametric versus parametric regression fits.
\newblock {\em The Annals of Statistics}, 21(4):1926--1947.

\bibitem[\protect\astroncite{H{\"a}rdle and Mammen}{1993b}]{hardle}
H{\"a}rdle, W. and Mammen, E. (1993b).
\newblock {Testing parametric versus nonparametric regression}.
\newblock {\em Annals of Statistics}, 21:1926--1947.

\bibitem[\protect\astroncite{Jain and Dubes}{1988}]{Jain88}
Jain, A.~K. and Dubes, R.~C. (1988).
\newblock {\em Algorithms for clustering data}.
\newblock Prentice-Hall, Inc., Upper Saddle River, NJ, USA.

\bibitem[\protect\astroncite{Kauermann and Opsomer}{2003}]{NoRefWorks:11}
Kauermann, G. and Opsomer, J. (2003).
\newblock {Local Likelihood Estimation in Generalized Additive Models}.
\newblock {\em Scandinavian Journal of Statistics}, 30:317--337.

\bibitem[\protect\astroncite{Kaufman and Rousseeuw}{1990}]{books/wi/KaufmanR90}
Kaufman, L. and Rousseeuw, P.~J. (1990).
\newblock {\em Finding Groups in Data: An Introduction to Cluster Analysis.}
\newblock John Wiley.

\bibitem[\protect\astroncite{Keinosuke and Hostetler}{1975}]{meanShift}
Keinosuke, F. and Hostetler, L.~D. (1975).
\newblock The estimation of the gradient of a density function, with
  applications in pattern recognition.
\newblock {\em IEEE Transactions on Information Theory}, 21(1):32--40.

\bibitem[\protect\astroncite{King et~al.}{1991}]{king91}
King, E., Hart, J.~D., and Wehrly, T.~E. (1991).
\newblock Testing the equality of two regression curves using linear smoothers.
\newblock {\em Statistics and Probability Letters}, 12(3):239--247.

\bibitem[\protect\astroncite{Kulasekera}{1995}]{kulasekera95}
Kulasekera, K.~B. (1995).
\newblock Comparison of regression curves using quasi-residuals.
\newblock {\em Journal of the American Statistical Association},
  90(431):1085--1093.

\bibitem[\protect\astroncite{Lang and Brezger}{2004}]{Lang04}
Lang, S. and Brezger, A. (2004).
\newblock Bayesian p-splines.
\newblock {\em Journal of Computational and Graphical Statistics}, 13:183--212.

\bibitem[\protect\astroncite{Liu}{1988}]{liu:1988}
Liu, R.~Y. (1988).
\newblock {Bootstrap Procedures under some Non-I.I.D. Models}.
\newblock {\em The Annals of Statistics}, 16(4):1696--1708.

\bibitem[\protect\astroncite{Macqueen}{1967}]{MacQueen67}
Macqueen, J.~B. (1967).
\newblock {\em Some methods of classification and analysis of multivariate
  observations}, volume~1.
\newblock Proceedings of the Fifth Berkeley Symposium on Mathematical
  Statistics and Probability (Univ. of Calif. Press).

\bibitem[\protect\astroncite{Mammen}{1993}]{citeulike:991599}
Mammen, E. (1993).
\newblock {Bootstrap and Wild Bootstrap for High Dimensional Linear Models}.
\newblock {\em The Annals of Statistics}, 21(1):255--285.

\bibitem[\protect\astroncite{Pardo-Fern{\'a}ndez et~al.}{2007}]{Pardo}
Pardo-Fern{\'a}ndez, J.~C., {Van Keilegom}, I., and Gonz{\'a}lez-Manteiga, W.
  (2007).
\newblock Testing for the equality of k regression curves.
\newblock {\em Statistica Sinica}, 17:1115--1137.

\bibitem[\protect\astroncite{Park et~al.}{2014}]{park2014}
Park, C., Hannig, J., and Kang, K.-H. (2014).
\newblock Nonparametric comparison of multiple regression curves in
  scale-space.
\newblock {\em Journal of Computational and Graphical Statistics},
  23(3):657--677.

\bibitem[\protect\astroncite{Park and Kang}{2008}]{park2008}
Park, C. and Kang, K.-H. (2008).
\newblock Sizer analysis for the comparison of regression curves.
\newblock {\em Computational Statistics and Data Analysis}, 52(8):3954--3970.

\bibitem[\protect\astroncite{Rosenblatt}{1975}]{Rosenblatt1975}
Rosenblatt, M. (1975).
\newblock A quadratic measure of deviation of two-dimensional density estimates
  and a test of independence.
\newblock {\em The Annals of Statistics}, 3(1):1--14.

\bibitem[\protect\astroncite{Ruppert et~al.}{1995}]{1995}
Ruppert, D., Sheather, S.~J., and Wand, M.~P. (1995).
\newblock An effective bandwidth selector for local least squares regression.
\newblock {\em Journal of the American Statistical Association},
  90(432):1257--1270.

\bibitem[\protect\astroncite{Sestelo and
  Roca-Pardi{\~n}as}{2019}]{sesteloJRSSC}
Sestelo, M. and Roca-Pardi{\~n}as, J. (2019).
\newblock {Testing critical points of non-parametric regression curves:
  application to the management of stalked barnacles}.
\newblock {\em Journal of the Royal Statistical Society C}, 68(4):1051--1070.

\bibitem[\protect\astroncite{Sestelo et~al.}{2017}]{sestelojss}
Sestelo, M., Villanueva, N.~M., Meira-Machado, L., and Roca-Pardi{\~n}as, J.
  (2017).
\newblock {npregfast: An R Package for Nonparametric Estimation and Inference
  in Life Sciences}.
\newblock {\em Journal of Statistical Software}, 82(12):1--27.

\bibitem[\protect\astroncite{Sestelo et~al.}{2016}]{npregfast}
Sestelo, M., Villanueva, N.~M., and Roca-Pardi{\~n}as, J. (2016).
\newblock {npregfast: Nonparametric Estimation of Regression Models with
  Factor-by-Curve Interactions}.
\newblock R package version 1.4.0.

\bibitem[\protect\astroncite{Srihera and
  Stute}{2010}]{Srihera:2010:NCR:1837522.1837642}
Srihera, R. and Stute, W. (2010).
\newblock Nonparametric comparison of regression functions.
\newblock {\em Journal of Multivariate Analysis}, 101:2039--2059.

\bibitem[\protect\astroncite{Tarpey}{2007}]{tarpey07}
Tarpey, T. (2007).
\newblock Linear transformations and the k-means clustering algorithm.
\newblock {\em The American Statistician}, 61(1):34--40.

\bibitem[\protect\astroncite{Villanueva et~al.}{2020}]{nora_clustcurv}
Villanueva, N.~M., , Sestelo, M., Meira-Machado, L., and Roca-Pardi{\~n}as, J.
  (2020).
\newblock clustcurv: An r package for determining groups in multiple curves.
\newblock {\em {Manuscript submitted for publication}}.

\bibitem[\protect\astroncite{Villanueva and Sestelo}{2020}]{clustcurv}
Villanueva, N.~M. and Sestelo, M. (2020).
\newblock {clustcurv: Determining Groups in Multiple Curves}.
\newblock R package version 2.0.1.

\bibitem[\protect\astroncite{Villanueva et~al.}{2019}]{villanueva2019}
Villanueva, N.~M., Sestelo, M., and Meira-Machado, L. (2019).
\newblock {A Method for Determining Groups in Multiple Survival Curves}.
\newblock {\em Statistics in Medicine}, 38:366--377.

\bibitem[\protect\astroncite{Vogt and Linton}{2017}]{vogt2016}
Vogt, M. and Linton, O. (2017).
\newblock Classification of non-parametric regression functions in longitudinal
  data models.
\newblock {\em Journal of the Royal Statistical Society Series B}, 79(1):5--27.

\bibitem[\protect\astroncite{Vogt and Linton}{2020}]{vogt2020}
Vogt, M. and Linton, O. (2020).
\newblock Multiscale clustering of nonparametric regression curves.
\newblock {\em Journal of Econometrics}, 216(1):305--325.

\bibitem[\protect\astroncite{Wand and Jones}{1995}]{NoRefWorks:7}
Wand, M.~P. and Jones, M.~C. (1995).
\newblock {\em {Kernel Smoothing}}.
\newblock Chapman {\&} Hall: London.

\bibitem[\protect\astroncite{Wu}{1986}]{citeulike:785121}
Wu, C. F.~J. (1986).
\newblock {Jackknife, Bootstrap and other resampling methods in regression
  analysis}.
\newblock {\em The Annals of Statistics}, 14(4):1261--1295.

\bibitem[\protect\astroncite{Young and Bowman}{1995}]{bowman95}
Young, S.~G. and Bowman, A.~W. (1995).
\newblock Nonparametric analysis of covariance.
\newblock {\em Biometrics}, 51:920--931.

\end{thebibliography}






\end{document}